\documentclass[letter]{jpsj3}
\usepackage{txfonts}

\title{
A Study of Ni-Substitution Effects on Heavy-Fermion CeCu$_2$Si$_2$ \\
--- Similarities between Ni Substitution and High-Pressure Effects ---
}
\author{
\name{Yoichi \surname{Ikeda}} $^1$ $^{\thanks{E-mail address: yo1-iked@science.okayama-u.ac.jp}}$, 
\name{Shingo \surname{Araki}} $^1$, 
\name{Tatsuo C. \surname{Kobayashi}} $^1$, \\
\name{Yusei \surname{Shimizu}} $^2$, 
\name{Tatsuya \surname{Yanagisawa}} $^2$, 
and \name{Hiroshi \surname{Amitsuka}} $^2$
}
\inst{
\address{$^1$Department of Physics, Okayama University, Okayama 700-8530, Japan} \\
\address{$^2$Graduate School of Science, Hokkaido University, N10W8, Kita-ku, Sapporo 060-0810, Japan} \\
}
\abst{
The effects of Ni substitution on Ce(Cu$_{1-x}$Ni$_x$)$_2$Si$_2$ have been studied by specific heat and electrical resistivity measurements. 
The specific heat measurement has revealed that the enhanced quantum fluctuations around an antiferromagnetic quantum critical point are markedly suppressed by Ni substitution, 
and that the Fermi liquid state recovers in the Ni-rich region ($x \geq$ $\sim$ 0.12). 
The characteristic $T$-linear dependence of the resistivity has been observed at approximately $x \sim$ 0.10 together with a sign of superconductivity. 
The variation of $n$ in the form of $\mathit{\rho}$ $-$ $\mathit{\rho}_0 = aT^n$ against $T_1^{\mathrm{max}}$, at which the resistivity peaks, coincides with the case of high-pressure experiments on pure CeCu$_2$Si$_2$. 
The anomalous $T$-linear behavior appears to occur in the crossover region from the Kondo regime to the valence fluctuation regime rather than in the conventional antiferromagnetic quantum critical region. 
}

\kword{CeCu$_2$Si$_2$, Ni substitution, heavy fermion, valence fluctuation, $T$-linear resistivity}

\begin{document}
\maketitle
The heavy-fermion superconductor CeCu$_2$Si$_2$ has been a topic of great interest for more than three decades \cite{Steglich1979}. 
This material is considered to be located around a three-dimensional (3D) antiferromagnetic (AFM) quantum critical point (QCP) without any physical/chemical tuning \cite{Steglich1996, Gegenwart, Aliev1983}. 
The superconductivity (SC) of CeCu$_2$Si$_2$ has a line-node superconducting gap and competes with an incommensurate AFM order, which is called the A-phase, with highly reduced staggered moments \cite{Nakamura1989, Bruls1994, Stockert2004}. 
What is fascinating is that previous pressure experiments revealed a clear increase in the superconducting transition temperature $T_{\mathrm{c}}$ at high pressures \cite{Thomas1993, Holmes}, where the AFM spin fluctuations are considerably suppressed compared with that of ambient pressure \cite{Fujiwara2008}. 
Recent $^{63}$Cu nuclear quadrupole resonance (Cu-NQR) measurements have revealed that bulk SC with a line-node gap structure is certainly realized at high pressures and suddenly disappears above a critical pressure $p_{\textrm{c}}$ $\sim$ 4.5 GPa \cite{Fujiwara2008}. 
In addition, two clearly distinct superconducting domes in the $p$-$T$ phase diagram were found in a Ge-substituted system CeCu$_2$(Si$_{1-x}$Ge$_{x}$)$_2$ \cite{Yuan2000}. 
For the SC around ambient pressure, AFM spin fluctuations have been considered to play an important role in the formation of the Cooper pair as well as other Ce-based heavy-fermion superconductors such as CeIn$_3$ \cite{CePd2Si2}, CePd$_2$Si$_2$, \cite{CePd2Si2} and CeRhIn$_5$ \cite{CeRhIn5}. 
In the light of experimental results, the AFM spin fluctuation scenario for the unconventional SCs, however, does not seem to be applicable to the high-pressure SC of CeCu$_2$Si$_2$. 

The critical charge-, valence-, or orbital-fluctuation scenarios are presumable theoretical ideas that can be used to explain the high-pressure SC of CeCu$_2$Si$_2$ and other anomalous metallic states observed in Ce- and Yb-based heavy-fermion systems \cite{Miyake, Watanabe, HIkeda, Hattori}. 
In CeCu$_2$Si$_2$, the linear temperature variation of the electrical resistivity, $\mathit{\rho} - \mathit{\rho}_0$ $\sim$ $T$, the marked increase in the residual resistivity, $\mathit{\rho}_0$, and the sudden decrease in the $T^2$ coefficient of the resistivity, $A$, are observed around the critical pressure $p_{\mathrm{c}}$. 
In addition, a significant increase in the Ce valence, which means the delocalization of the $f$-electron on the Ce ion, is manifested according to the latest studies of the Cu-NQR \cite{NQR} and the Ce-$L_{\mathrm{III}}$ edge X-ray absorption spectra (XAS) at high pressures \cite{Rueff2011}. 
These characteristics have been believed to be signs of the critical behavior associated with a valence transition around the critical end-point \cite{Holmes, Miyake, Watanabe}, while we have been requiring, in the course of our studies \cite{Araki}, more signatures for the clarification of fascinating quantum phenomena. 

We have reviewed prior research studies for a pseudo-ternary Ce(Cu$_{1-x}$Ni$_x$)$_2$Si$_2$ system, which were reported by two groups about 25 years ago \cite{Sampathkumaran1986-1987, Edwards1987}, and found it interesting to reinvestigate the series from the point of view of further understanding the quantum criticality in CeCu$_2$Si$_2$. 
The counterpart material CeNi$_2$Si$_2$ is the so-called intermediate valence compound with a relatively high Kondo temperature, $T_{\mathrm{K}}$, on the order of several hundred K \cite{Neifeld 1985}, such that Ni substitution greatly transforms a heavy-fermion state into a mixed-valence state as mentioned in the previous reports. 
The rapid evolution of the ground state resembles the case of high-pressure experiments on CeCu$_2$Si$_2$; therefore, we expect similar quantum phenomena in the Ni-substituted system as well. 

Polycrystalline samples were grown by arc melting and annealed under Cu atmosphere in a quartz tube at 1100 $-$ 1200 $^{\circ}$C for one week. 
All samples crystallized in the ThCr$_2$Si$_2$-type structure were characterized by powder X-ray diffraction (XRD) analysis. 
The Ce$_2$CuSi$_3$ phase was mixed in the as-grown samples, but fully diminished in the annealed ones within the experimental sensitivity of XRD measurements as reported previously \cite{Sun1990}. 
Annealed-sample surfaces were polished to remove a small number of Cu-rich binary-impurity phases before measurements. 
Electrical resistivity measurements were carried out by an AC/DC four-terminal technique between 0.1 and 300 K. 
Specific heat was measured by a thermal relaxation method in Physical Property Measurement System (Quantum Design Inc.) between 2 and 30 K. 

The lattice constants of Ce(Cu$_{1-x}$Ni$_x$)$_2$Si$_2$ at room temperature are shown in Fig. 1(a). 
Both $a$- and $c$-axes (lattice constants) monotonically decrease with increasing Ni concentration $x$.
In this paper, $x$ means the initial composition. 
Note that the difference in initial composition is about $\pm$ 1\% according to the results of chemical composition analysis except for $x$ = 0.13, which is comparable to $x$ = 0.11 $-$ 0.12. 
The $c$-axis lattice constant shows a major change compared with the $a$-axis lattice constant. 
The lattice volume [Fig. 1(b)] varies in a linear manner (Vegard's law) for the entire $x$ range without any change in lattice structure or a marked jump, such as a first-order valence transition observed in other rare-earth (Eu, Yb) compounds \cite{Eu, Yb}. 
The relative change in the volume from $x$ = 0.0 to 0.3 is about $-2$\%. 
This volume contraction corresponds to the pressurization of about 3 $\sim$ 5 GPa according to the bulk compressibility of CeCu$_2$Si$_2$ at room temperature \cite{Spain1986}.  

Results of the specific heat measurement are shown in Fig. \ref{f5}, where $C_{4f}$ values are obtained by subtracting nonmagnetic contributions from raw data. 
In this study, the specific heat of LaCu$_2$Si$_2$ with the Sommerfeld coefficient $\mathit{\gamma}$ $\sim$ 5.2 mJ$\cdot$K$^{-2}\cdot$mol$^{-1}$ and the Debye temperature $\mathit{\Theta}_{\mathrm{D}}$ $\sim$ 410 K is used as the nonmagnetic reference. 
We consider the low-temperature upturn of $C_{4f}$/$T$ in light of the AFM quantum criticality as discussed by Gegenwart \textit{et al} \cite{Gegenwart}. 
In a limited temperature range below $T$ $\sim$ 4 K, we can roughly fit $C_{4f}$/$T$ in the form of $\mathit{\gamma} - \beta T^{0.5}$, where $\mathit{\gamma}$ denotes an electronic specific heat coefficient and the second term is a correction term for $T$-dependent $\mathit{\gamma}$ in the case of a 3D AFM QCP of an itinerant magnet \cite{Moriya}. 
The estimated values are shown in the inset of Fig. \ref{f5}. 
Enhancements of $\mathit{\gamma}$ and $\mathit{\beta}$, \textit{i.e.,} the low-temperature upturn of $C_{4f}$/$T$, are suppressed by increasing the Ni concentration. 
Therefore, the Ni substitution appears to suppress the critical low-lying AFM spin fluctuations. 
Furthermore, it seems that the low-lying fluctuations do not remain in the Ni-rich region ($x \geq$ 0.12), because $C_{4f}$/$T$ becomes constant ($\mathit{\beta}$ $\sim$ 0) at low temperatures. 
This temperature variation ($C_{4f}$/$T$ = constant) indicates that the Fermi liquid state is fully recovered above $x$ $\sim$ 0.12. 
This finding can be confirmed by the relationship between $\mathit{\gamma}$ and the Pauli paramagnetic susceptibility $\mathit{\chi}_0$: $\mathit{\gamma}$ and $\mathit{\chi}_0$ of the series ($x \geq \sim$ 0.12) are located around a universal line with the Wilson ratio $R_{\mathrm{W}}$ = 2 along with various Ce-based intermediate valence compounds (not shown) \cite{KW}. 
In addition, the suppression of $\mathit{\gamma}$ by Ni substitution is similar to the results of Cu-NQR measurement under a high pressure in pure CeCu$_2$Si$_2$. \cite{Fujiwara2008}

The use of the Bethe ansatz formula of the doublet ground state, $T_{\textrm{K}}$ = $w \pi R/6\mathit{\gamma}$ \cite{Rajan}, where $w$ = 1.294 is the Wilson number and $R$ is the gas constant, gives $T_{\mathrm{K}}$ for Ce(Cu$_{1-x}$Ni$_x$)$_2$Si$_2$, as plotted in Fig. \ref{f3}. 
For $x$ = 0.0, $T_{\mathrm{K}}$ is estimated to be $\sim$ 7 K, which is in good agreement with a previous NMR result \cite{TK}.
In particular, it is interesting to note that a rapid increase in $T_{\mathrm{K}}$ is observed for $x$ = 0.09 $-$ 0.12, in which the low-temperature enhancement of $C_{4f}$/$T$ almost completely disappears. 
The precipitous increase in $T_{\mathrm{K}}$ suggests that the transformation from the Kondo (heavy-fermion) regime with $n_f$ $\sim$ 1, where $n_f$ is the $f$-electron number per Ce ion, to the valence fluctuation regime with $n_f$ $<$ 1 occurs in the Ni-substituted system, as mentioned in the previous studies \cite{Sampathkumaran1986-1987, Edwards1987}. 
In fact, we can restate this as follows: a valence crossover is realized in the Ni-rich region ($x \geq$ $\sim 0.12$), although the crossover is not sharp compared with the $\alpha-\gamma$ transition in the Ce metal \cite{Ce} and the valence transition with a major change in $n_f$ in the Eu and Yb compounds \cite{Eu, Yb}. 
This state is consistent with the results of the previous Ce-$L_{\mathrm{III}}$ XAS experiments on the Ni-substituted system \cite{Edwards1987}, which show a gradual valence change in a Ni-rich region.

Results of electrical resistivity measurement are shown in the inset of Fig. \ref{f3}. 
Since the resistivity in the present measurement varied even in the same sample owing to the possible scattering by microcracks, we were unable to analyze the absolute value of the resistivity precisely. 
The resistivity for $x$ = 0.0 shows two peaks due to the Kondo and crystalline electric field (CEF) effects in the Kondo-lattice system \cite{Yamada}, where the peak temperatures are defined as $T_1^{\mathrm{max}}$ and $T_2^{\mathrm{max}}$ ($T_1^{\mathrm{max}}$ $\leq$ $T_2^{\mathrm{max}}$) \cite{Holmes}. 
As the Ni concentration increases, $T_1^{\mathrm{max}}$ increases more rapidly than $T_2^{\mathrm{max}}$, which merges the resistivity peaks above $x \sim$ 0.10, as shown in Fig. \ref{f3}. 
This tendency qualitatively coincides with those previously reported \cite{Sampathkumaran1986-1987, Edwards1987}, although the two peaks merge at slightly lower Ni concentrations in our results. 
The Ni doping dependence of the two characteristic temperatures indicates that $T_{\mathrm{K}}$ approaches the CEF splitting energy.  
The effects of Ni substitution on $T_1^{\mathrm{max}}$ and $T_2^{\mathrm{max}}$ closely resemble the results of high-pressure experiments on pure CeCu$_2$Si$_2$ and CeCu$_2$Ge$_2$ \cite{Holmes, Jaccard}. 
In the case of CeCu$_2$Si$_2$ in a hydrostatic pressure, $T_1^{\mathrm{max}}$ and $T_2^{\mathrm{max}}$ are distinguishable from each other at 3 $-$ 5 GPa. 

Figure \ref{f2} shows the temperature dependence of electrical resistivity, which is shifted by a factor of $\mathit{\rho}_{\mathrm{offset}}$, below 5 K at various Ni concentrations. 
An apparent superconducting transition is observed at $x$ = 0.0 ($T_{\mathrm{c}}$ = 0.67 K) and 0.01 (0.59 K). 
The low-temperature behavior of $x$ = 0.0 is the same as that of S-type CeCu$_2$Si$_2$ \cite{Gegenwart}, in which specific heat shows no magnetic transitions in zero field (Fig. \ref{f5}). 
The SC is suppressed by Ni substitution and completely disappears at $x$ = 0.05 and 0.09. 
However, fractional resistivity drops are observed in Ni-rich ($x \geq$ 0.10) compounds. 
Although the zero resistance because of a pair-breaking effect has not been observed, the resistivity drop is strikingly absent at $x$ = 0.05 and 0.09 but is obviously finite around $x$ = 0.10. 
The size of the resistivity drop due to the filamentary SC reaches a maximum around $x \sim$ 0.11 but decreases at high Ni concentrations. 
Therefore, it seems that the SC is recovered around $x$ $\sim$ 0.10. 
Furthermore, it is interesting to note that the filamentary SC appears to occur together with a characteristic $T$-linear dependence above $T_{\mathrm{c}}$. 

At low temperatures above $T_{\mathrm{c}}$, the resistivity can be fitted by a power law in the form of $\mathit{\rho} - \mathit{\rho}_0$ = $A'T^n$, where the temperature exponent $n$ is a fitting parameter. 
The temperature exponent $n$ of Ce(Cu$_{1-x}$Ni$_x$)$_2$Si$_2$ series is shown in Fig. \ref{f4} as a function of $T_1^{\mathrm{max}}$ with the results of high-pressure experiments \cite{Holmes}. 
The estimated $n$ is scattered around $n \sim$ 1.5 below $x$ = 0.09 ($T_1^{\mathrm{max}}$ $<$ 60 K). 
The $T^{\sim 1.5}$ dependence at $x$ = 0.05 and 0.09 is observed down to $\sim$ 0.1 K without an SC transition. 
The exponent ($n$ = 1.5) of $x$ = 0.05 and 0.09 coincides with the expected value for the 3D AFM quantum criticality of an itinerant magnet \cite{Moriya}, which is consistent with the results of specific heat measurement. 

Subsequently, the exponent is $n$ $\sim$ 1 around $x$ $\sim$ 0.10 ($T_1^{\mathrm{max}}$ = 60 $\sim$ 80 K), as shown in Fig. \ref{f2} and in the inset of Fig. \ref{f4}. 
It seems that the $T^{\sim 1.5}$ dependence changes into a nearly $T$-linear dependence around the same Ni concentration as that when the low-temperature enhancement of $C_{4f}$/$T$ and the two peaks in $\mathit{\rho}(T)$ disappear, as mentioned above. 
In Ni-rich compounds ($T_1^{\mathrm{max}}$ $\geq$ 100 K), the resistivity shows a quadratic temperature variation ($\mathit{\rho} - \mathit{\rho}_0$ = $AT^2$) with the recovery of the Fermi liquid state. 
The characteristic $T$-linear dependence has also been observed in pure CeCu$_2$Si$_2$ around $p_{\mathrm{c}}$, at which $T_{\mathrm{c}}$ becomes a maximum. 
The variation in $n$ against $T_1^{\mathrm{max}}$ coincides with the results of the high-pressure experiments, \cite{Holmes} as shown in Fig. \ref{f4}. 
Considering our specific heat results, 
it seems that the anomalous Fermi liquid behavior in $\mathit{\rho}(T)$ and the SC around $x$ $\sim$ 0.10 are affected by the disappearance of the low-lying antiferromagnetic fluctuations and/or the rapid increase in $T_{\mathrm{K}}$ (valence crossover). 
Intriguingly, the chemical trend for the SC in the Ni-doped system is similar to the cases of high-pressure experiments on CeCu$_2$Si$_2$ and the Ge-substituted system \cite{Yuan2000}: 
\textit{i.e.}, $T_{\mathrm{c}}$ shows two domes in the $p$-$T$ phase diagram. 
However, the reason why $T_{\mathrm{c}}$ is not so high compared with those of the Ge-substituted system around $x$ $\sim$ 0.12 is still unknown. 
The quantitative differences may arise from the changes in Fermi surface properties introduced by Ni substitution. 
According to a recent theoretical study, which suggests that an anisotropic SC is robust against nonmagnetic impurity scattering if critical valence fluctuations are developed \cite{robustSC}, the $T$-linear resistivity and the recovery of the SC in Ce(Cu$_{1-x}$Ni$_x$)$_2$Si$_2$ are possibly induced by critical valence fluctuations, although direct evidence is absent.

In conclusion, we studied the effects of the Ni substitution on CeCu$_2$Si$_2$ to investigate its attractive quantum phenomena. 
Specific heat measurements revealed that Ni-substitution suppresses the low-temperature enhancement of $C_{4f}/T$ due to quantum fluctuations around a 3D AFM QCP. 
The enhancement seems to disappear above $x \sim$ 0.12, at which $C_{4f}/T$ shows Fermi liquid behavior. 
Around the same Ni concentration, $T_{\mathrm{K}}$ becomes increasingly enlarged as the result of a rapid decrease in $\mathit{\gamma}$, and the two characteristic temperatures $T_1^{\mathrm{max}}$ and $T_2^{\mathrm{max}}$ in the resistivity become indistinguishable from each other. 
These results indicate that $T_{\mathrm{K}}$ approaches the CEF splitting energy. 
Furthermore, resistivity measurements clarified that the variation of the temperature exponent $n$, which is the results of a fit to $\mathit{\rho} - \mathit{\rho}_0$ = $A'T^n$ at low temperatures, against $T_1^{\mathrm{max}}$ coincides with the results of high-pressure experiments on pure CeCu$_2$Si$_2$. 
Intriguingly, a nearly $T$-linear dependence, together with a filamentary SC, is observed around $x$ $\sim$ 0.10. 
The anomalous Fermi liquid behavior and the recovery of the SC around $x \sim$ 0.10, and perhaps the case of pure CeCu$_2$Si$_2$ at high pressures, seem to occur in the crossover region from the Kondo regime to the valence fluctuation regime rather than in the conventional AFM quantum critical region. 

\smallskip
\begin{acknowledgments}
We would like to thank Professor K. Miyake and Professor K. Fujiwara for helpful discussions and comments. 
The author Y. I. also acknowledges Professor Y. Kubozono, Professor M. Nohara, Professor T. Kambe, Dr. K. Kudo, and Mr. H. Urakami for a great deal of support in cryogenic experiments. 
\end{acknowledgments}

\begin{figure}[b]
\begin{center}
\includegraphics[clip, width=7cm, height=7cm]{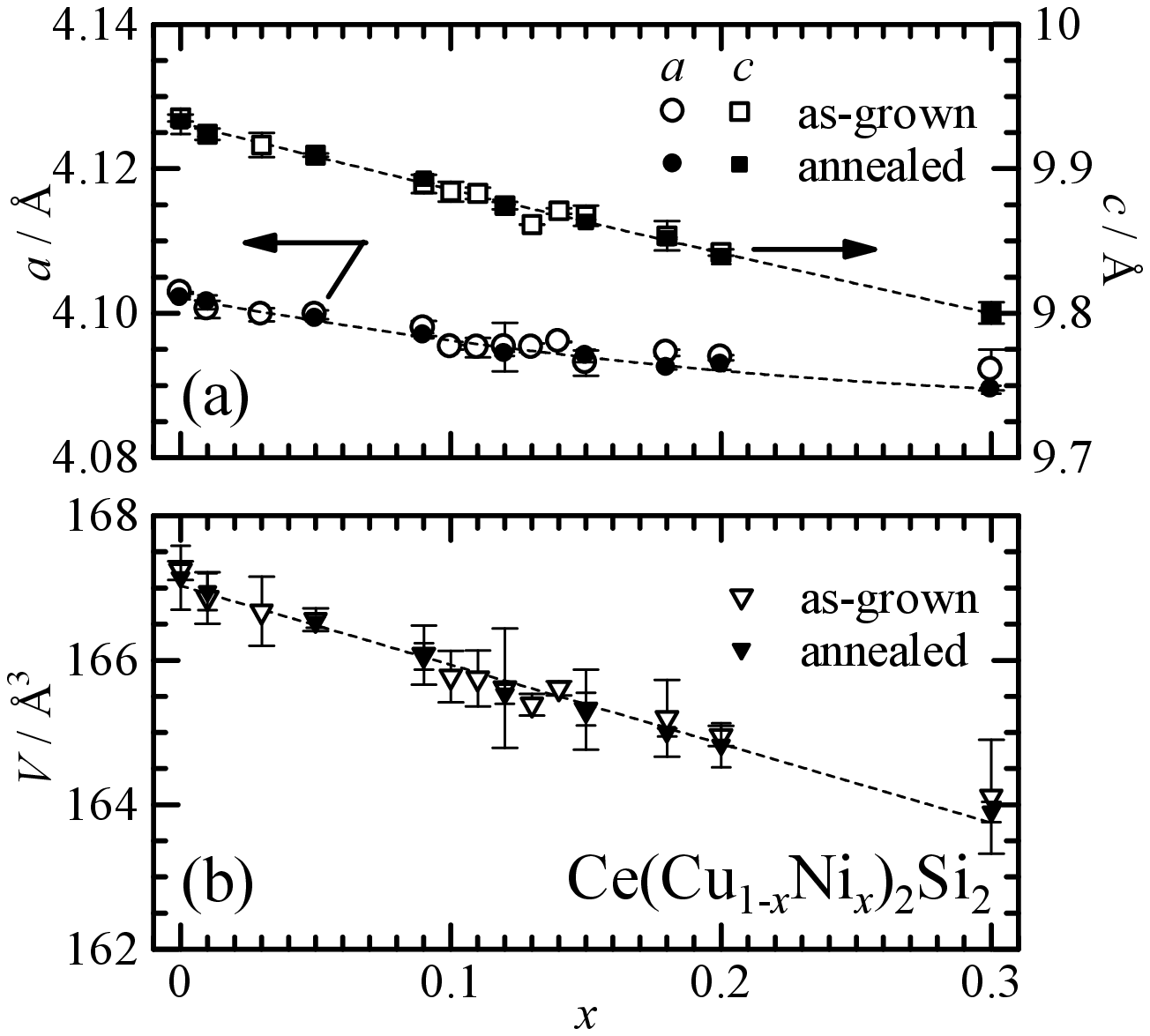}
\caption{(a) Lattice constants and (b) volumes of Ce(Cu$_{1-x}$Ni$_x$)$_2$Si$_2$ at room temperature as a function of Ni concentration $x$. Open and closed symbols correspond to as-grown and annealed samples, respectively. Dashed lines serve as visual guides.}
\label{f1}
\end{center}
\end{figure}

\begin{figure}[]
\begin{center}
\includegraphics[clip, width=7cm, height=7cm]{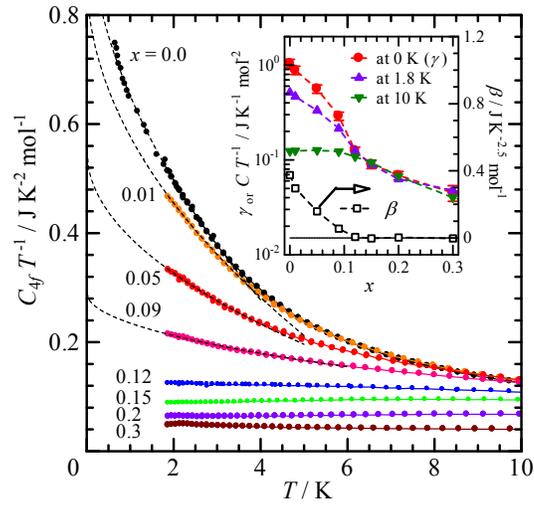}
\caption{
(Color online) $C_{4f}$/$T$ of the Ce(Cu$_{1-x}$Ni$_x$)$_2$Si$_2$ series as a function of temperature, where $C_{4f}$ = $C$($T$) - $C$(LaCu$_2$Si$_2$). 
Dashed lines represent the result of a fit to $C_{4f}$/$T$ = $\mathit{\gamma} - \beta T^{0.5}$ below $T$ $\sim$ 4 K. 
Inset: $C_{4f}$/$T$ ($\mathit{\gamma}$) and $\mathit{\beta}$ versus Ni concentration $x$. }
\label{f5}
\end{center}
\end{figure}

\begin{figure}
\begin{center}
\includegraphics[clip, width=7cm, height=7cm]{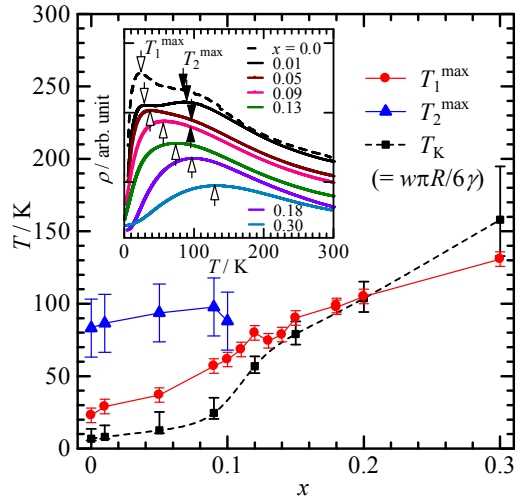}
\end{center}
\caption{(Color online) Characteristic temperatures ($T_1^{\mathrm{max}}$, $T_2^{\mathrm{max}}$, and $T_{\mathrm{K}}$) versus Ni concentration $x$. Precise definitions of these temperatures are written in the text. 
Solid and dashed lines serve as visual guides. 
The inset shows the temperature dependence of the normalized resistivity at various Ni concentrations. The open and closed arrows show $T_1^{\mathrm{max}}$ and $T_2^{\mathrm{max}}$, respectively.}
\label{f3}
\end{figure}

\begin{figure}
\begin{center}
\includegraphics[clip, width=7cm, height=7cm]{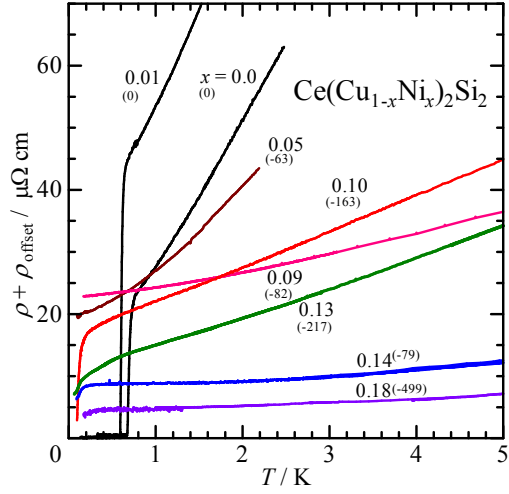}
\end{center}
\caption{(Color online) Temperature dependence of the electrical resistivity of Ce(Cu$_{1-x}$Ni$_x$)$_2$Si$_2$. 
The curves for $x$ = 0.05, 0.09, 0.10, 0.13, 0.14, and 0.18 have been vertically shifted for display. 
The small numbers in parentheses correspond to the offset values $\mathit{\rho}_{\mathrm{offset}}$. 
}
\label{f2}
\end{figure}

\begin{figure}
\begin{center}
\includegraphics[clip, width=7.0cm, height=7.0cm]{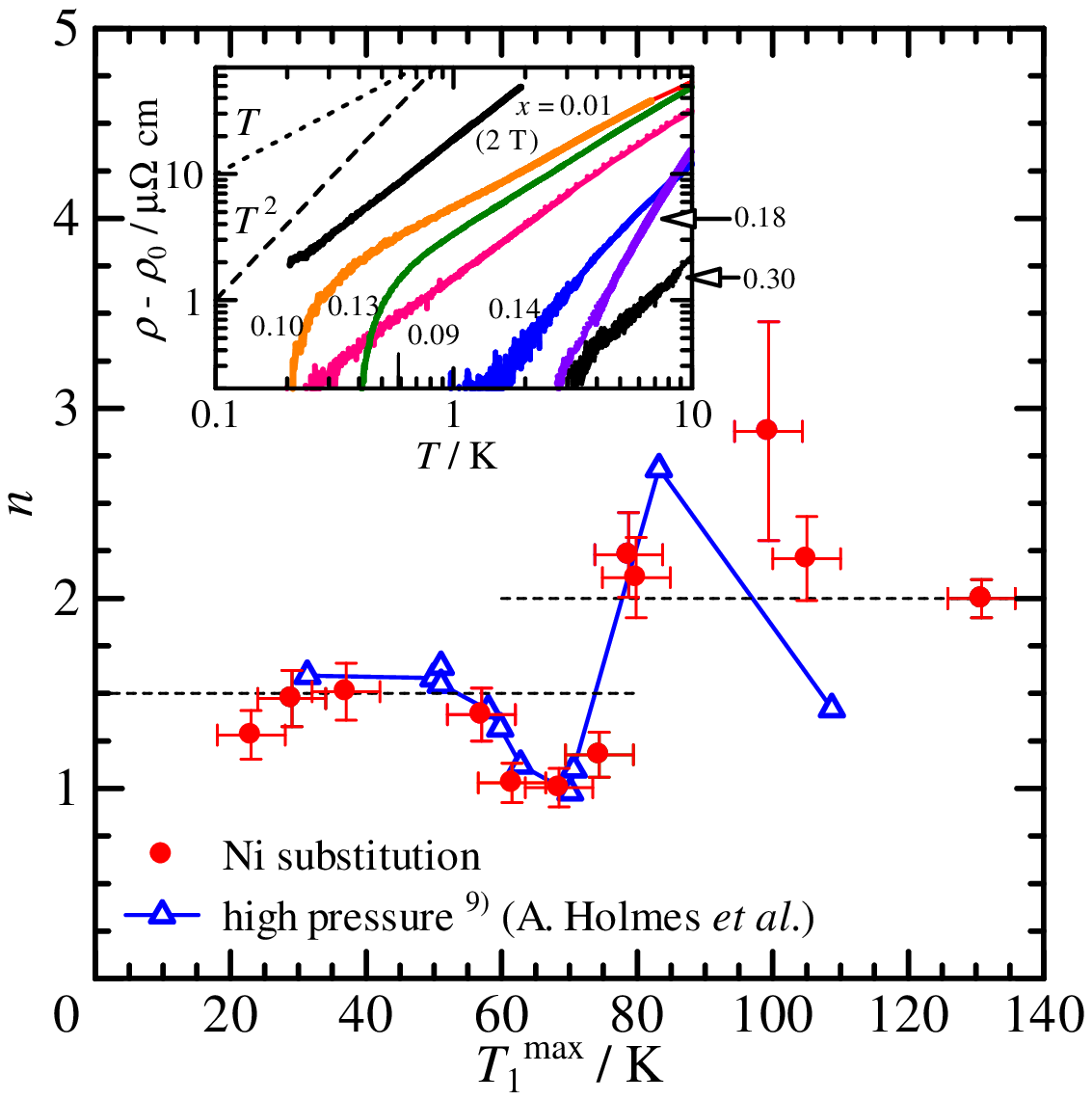}
\end{center}
\caption{(Color online) Variation in temperature exponent $n$ in the form of $\mathit{\rho} - \mathit{\rho}_0$ = $A'T^n$ against $T_1^{\mathrm{max}}$ for the Ni-substituted system (circles) and the high-pressure experiments (triangles denote data taken from ref. 9). 
The inset shows the temperature dependence of $\mathit{\rho} - \mathit{\rho}_0$ in a log-log scale.}
\label{f4}
\end{figure}


\begin{thebibliography}{37}
\bibitem{KW} Although the relationship between  $\mathit{\gamma}$ and $A$ values, \textit{i.e.} the Kadowaki-Woods relationship, is also expected to follow a universal line, the same as that between $\mathit{\gamma}$ and $\mathit{\chi}_0$, unfortunately we could not determine the precise relationship between $A$ and $\mathit{\gamma}$.  
\end{thebibliography}
\end{document}